\documentclass[proceedings]{JHEP}
\conference{Integrable Theories, Solitons and
Duality, UNESP2002}
\usepackage{epsfig}
\def\bea{\begin{eqnarray}}
\def\eea{\end{eqnarray}}
\def\EQ{\begin{equation}}
\def\EN{\end{equation}}

\title{Universal Ratios and Correlation Functions\footnote{
Talk presented by G. Mussardo.}}

\author{G. Delfino and G. Mussardo \\Scuola Internazionale 
Superiore di Studi Avanzati (SISSA) \\
INFN, Sezione di Trieste (Italy)} 
\abstract{We review some recent results concerning the quantitative 
analysis of the universality classes of two-dimensional statistical 
models near their critical point. We also discuss the exact calculation 
of the two--point correlation functions of disorder operators in a free theory 
of complex bosonic and fermionic field, correlators ruled by a Painlev\`{e} 
differential equation.}

\begin{document}
\section{Introduction}

%{\large {\bf Introduction}}

\noindent
One of the most important successes of Quantum Field Theory (QFT) 
in recent years is the quantitative analysis of the
universality classes of two--dimensional statistical mechanical 
models near their second order phase transition points. 
This has been possible, in particular, thanks to new developments 
in the methods of computing off--critical correlation functions. 
The aim of this talk is to review some of the recent results relative 
to these two important subjects. Our discussion is divided in two parts: 
in the first part, we shall see how correlation functions can be used to 
extract useful numbers (the so--called {\em universal ratios}) that 
can be directly compared with experiments. In the second part, we 
shall concentrate on a particularly interesting class of models, those of 
fermionic and bosonic theories with a $Z_n$ symmetry, for which 
there is an elegant approach to derive differential equations 
satisfied by the two--point functions of their disorder 
operators. 

\section{Universal Amplitude Ratios}
%{\large {\bf Universal Amplitude Ratios}}

\noindent
Consider a statistical model with $n$ relevant fields 
$\varphi_i(x)$ at criticality. Near the critical point, 
its action can be parameterised as 
\EQ
{\cal A} = {\cal A}_{CFT} + g_i \int \varphi_i(x) d^Dx
\,\,\,, 
\label{actionoff}
\EN 
where ${\cal A}_{CFT}$ is the action which encodes the 
data of the Conformal Field Theory of the critical point. 
The short--distance behaviour of the two--point functions is given by 
\[ 
\langle \varphi_i (x) \varphi_i(0) \rangle
\simeq \frac{1}{\hspace{1mm}|x|^{4 \Delta_i}}
\]
for $|x|\rightarrow 0$, where $\Delta_i$ is the conformal 
dimension of the operator $\varphi_i$. Hence, their conjugate 
coupling constants $g_i$ 
behave as
$ 
g_i \sim \Lambda ^{D - 2 \Delta_i} \,\,\, ,
$ where $\Lambda$ is a mass scale. Away from criticality 
there will be generally a finite correlation length $\xi$ which in the 
thermodynamical limit scales as
$
\xi \sim a \,(K_i g_i)^{-\frac{1}{D - 2 \Delta_i}} \,\,\,,
$
where $a \sim \Lambda^{-1}$ may be regarded as a microscopic
length scale. The dimensionless quantities $K_i$ are non--universal 
metric factors specific of the system representing the universality class
and depending on the units chosen for measuring the external sources
$g_i$. 
If there are several deformations of the conformal action, the most 
general expression for the scaling form of the correlation length 
may be written as
\EQ
\xi =\xi_i \equiv a \,(K_i g_i)^
{-\frac{1}{D -2 \Delta_i}}\,
{\cal L}_i\left(\frac{K_j g_j}{(K_i g_i)^{\phi_{ji}}}\right) \,\,\,,
\label{xii}
\EN
where
$ 
\phi_{ji} \equiv \frac{D-2 \Delta_j}{D-2 \Delta_i}
$
are the so--called {\em crossover exponents} whereas ${\cal L}_i$        
are universal homogeneous scaling functions of the ratios
$\frac{K_j g_j}{(K_i g_i)^{\phi_{ji}}}$. There are of course
several (but equivalent) ways of expressing these scaling forms,
depending on which coupling constant is selected as a prefactor.
In the limit where $g_l \rightarrow 0$ ($l \neq i$) but
$g_i \neq 0$, equation (\ref{xii}) becomes
\EQ
\xi_i = a \,\xi_i^0 \,g_i
^{-\frac{1}{D -2 \Delta_i}}
\,\,\,\,, \,\,\,\, 
\xi_i^0 \sim K_i^{-\frac{1}{D -2 \Delta_i}} \,\,\,.
\label{xio}
\EN
Let us consider now the free--energy $\hat f[g_1,\ldots,g_n]$, 
a dimensionless quantity defined by
\EQ
e^{-\hat f(g_1,\ldots,g_n)} = 
\int {\cal D}\phi \, e^{- \left[
{\cal A}_{CFT} +\sum_{i=1}^n g_i \int \varphi_i(x) d^Dx
\right]} \,\,\,.
\label{free}
\EN
Under the hypothesis of hyperscaling, its singular part (per unit 
of volume) $f[g_1,\ldots,g_n]$ is proportional, in the thermodynamical 
limit, to the $D$-th power of the correlation length. Depending on the 
scaling form adopted for the correlation length, we have correspondingly
several (but equivalent) ways of parameterizing this quantity
\EQ
f_i[g_1,\ldots,g_n] \equiv
\left(K_i g_i\right)^{\frac{D}{D-2\Delta_i}} \,
{\cal F}_i\left(\frac{K_j g_j}{(K_i g_i)^{\phi_{ji}}}\right)
\,\,\,.
\label{scalingfree}
\EN
The functions ${\cal F}_i$ are universal homogeneous scaling
functions of the ratios $\frac{K_j g_j}{(K_i g_i)^{\phi_{ji}}}$.

From the above expression we can derive the expectation values of several 
quantities. We will use the notation $\langle ... \rangle_i$ to denote 
expectation values computed in the off--critical theory obtained by 
keeping (at the end) only the coupling constant $g_i$ different from zero.
Basic quantities are the vacuum expectation values (VEV) of the fields 
$\varphi_j$ which can be parameterized as
\EQ
\langle \varphi_j \rangle_i = -\left.\frac{\partial f_i}{\partial g_j}
\right|_{g_l=0} \equiv B_{ji}
g_i^{\frac{2 \Delta_j}{D-2 \Delta_i}} \,\,\,,
\label{vacuumji}
\EN
with
\EQ
B_{ji} \sim K_j K_i^{\frac{2 \Delta_j}{D-2 \Delta_i}} \,\,\,.
\label{bji}
\EN
The above relations can be equivalently expressed as
\EQ
g_i = D_{ij} \left(\langle \varphi_j\rangle_i\right)^
{\frac{D-2 \Delta_i}{2 \Delta_j}} \,\,\,,
\label{magn}
\EN
with
\EQ
D_{ij} \sim \frac{1}{K_i K_j^{\frac{D-2 \Delta_i}{2 \Delta_j}}}
\,\,\,.
\label{dij}
\EN
The generalized susceptibilities of the model are defined by
\EQ
\hat \Gamma_{jk}^i =
\frac{\partial}{\partial g_k} \langle \varphi_j\rangle_i =\left.
-\frac{\partial^2 f_i}{\partial g_k \partial g_j}\right|_{g_l=0}
\,\,\,. \label{susc}
\EN
They are obviously symmetrical in the two lower indices. By
extracting their dependence on the coupling constant $g_i$,
they can be expressed as
\EQ
\hat \Gamma_{jk}^i = \Gamma_{jk}^i \,g_i^
{\frac{2 \Delta_j + 2 \Delta_k - D}
{D - 2 \Delta_i}} \,\,\, ,
\label{hatgammajki}
\EN
with
\EQ
\Gamma_{jk}^i \sim K_j K_k K_i^{\frac{2 \Delta_j + 2 \Delta_k - D}
{D - 2 \Delta_i}} \,\,\, .
\label{gammajki}
\EN
The various quantities obtained by taking the derivatives of the 
free--energy obviously contain the quantities $K_i$ which make their 
values not universal. However, it is easy and always possible to 
consider special combinations thereof in such a way to cancel out all 
metric factors. Universal Ratios usually considered in the literature 
are those given below 
\EQ
(R_c)^i_{jk} = \frac{\Gamma_{ii}^i \Gamma_{jk}^i}{B_{ji} B_{ki}}
\,\,\, ;
\label{Rc}
\EN
\EQ
(R_{\chi})^i_j = \Gamma_{jj}^i D_{jj} B_{ji}^
{\frac{D-4 \Delta_j}{2 \Delta_j}}
\,\,\,;
\label{Rchi}
\EN
\EQ
R^i_{\xi} = \left(\Gamma_{ii}^i\right)^{1/D} \xi_i^0 \,\,\,;
\label{Rxi}
\EN
\EQ
(R_A)^i_j = \Gamma_{jj}^i \, D_{ii}^
{\frac{4 \Delta_j + 2 \Delta_i - 2 D}{D-2 \Delta_i}} \,
B_{ij}^{\frac{2\Delta_j -D}{\Delta_i}} \,\,\, ;
\label{RA}
\EN
\EQ
(Q_2)^i_{jk} = \frac{\Gamma^i_{jj}}{\Gamma^k_{jj}}
\left(\frac{\xi_k^0}{\xi_j^0}\right)^{D-4 \Delta_j} \,\,\,.
\label{Q2}
\EN
From their definition, these quantities are pure numbers attached 
to the universality classes and therefore they can be used to characterize 
them. Contrary to the critical exponents (which are characteristic of 
the critical point) they carry information about the {\it scaling region}.
Moreover, the amplitude ratios are numbers which typically present
significant variations between different classes of universality, 
whereas the critical exponents usually assume small values which 
only vary by a small percent when changing the universality class. 
Hence the universal ratios may be ideal marks of the 
critical scaling regime \cite{Privman}. It is also worth emphasizing 
that, from an experimental point of view, it should be simpler to 
measure universal amplitude ratios rather than critical exponents: 
to determine the former quantities one needs to perform 
several measurements at a single, fixed value of the coupling 
which drives the system away from criticality whereas to determine 
the latter, one needs to make measurements over several decades 
along the axes of the off--critical couplings. Moreover, 
although not all of them are independent, the universal ratios 
are a larger set of numbers than the critical exponents and 
therefore permit a more precise determination of the class of
universality.

\section{Quantum Field Theory Approach}

It is interesting to see how in two dimensions Quantum Field Theory 
provides a powerful approach to compute the Universal Ratios of a 
given class of universality in a non-perturbative way. 
In the following we concentrate mostly 
our attention on the off--critical models obtained as deformations 
of the so--called {\em Minimal Models} of two--dimensional Conformal 
Field Theory \cite{BPZ,FQS1}.

As we have seen before, each coupling constant $g_i$ relative 
to the relevant operator $\varphi_i(x)$ of the model under 
investigation is a dimensional quantity which can be related to 
the lowest mass--gap $m_i = \xi_i^{-1}$ of the off--critical theory 
according to the formula 
\EQ
m_i = {\cal C}_i \,g_i^{\frac{1}{2-2 \Delta_i}} \,\,\, .  
\label{mg}
\EN 
When the QFT associated to the action (\ref{actionoff}) 
is integrable, the pure number ${\cal C}_i$ can be exactly 
determined by means of the Thermodynamical Bethe Ansatz 
\cite{TBA,fateev}. When the theory is not integrable, the constant 
${\cal C}_i$ can be nevertheless determined by a numerical method, 
based on the so--called Truncated Conformal Space Approach 
\cite{YZ}. In conclusion, for all individual deformations of 
a given model, we are able to completely set the relationship which 
links the coupling constant to the mass--gap of the theory and 
therefore to switch freely between these two variables. 

Additional quantities which can be determined  by QFT are the matrix 
elements of the order parameters, the simplest ones being the vacuum 
expectation values (VEV). They are parameterised as 
\EQ
\langle \varphi_j \rangle_i = B_{ji} \,g_i^{\frac{\Delta_j}
{1- \Delta_i}} 
\,\,\,.
\label{VEVg}
\EN 
When the theory is integrable, the constant $B_{ji}$ can 
be fixed exactly, thanks to the results of a remarkable series of 
papers \cite{russian1,russian2}. When it is not integrable, the 
constant $B_{ji}$ can be nevertheless estimated by means 
of a numerical approach, as firstly shown in \cite{GM1}. 
Hence, we are able to determine completely also these quantities. 
Moreover, as shown in \cite{FMS2}, a generalization of the 
numerical approach of ref.\,\cite{GM1} often leads to a reasonable 
estimate of the matrix elements of the order parameters between 
the vacuum states and some of the excited states, as for instance 
$
\langle 0 | \varphi_j | A_k \rangle_i 
$ 
where $A_k$ is a one--particle state of mass $M_k$. These 
quantities turn out to be useful for obtaining sensible
approximation of the large--distance behavior of several 
correlators. 

Another important and useful piece of information on the off--critical 
models can be obtained by exploiting the properties of the stress--energy 
tensor $T_{\mu \nu}(x)$. In the presence of the perturbing field 
$\varphi_i$, the trace of the stress--energy tensor is 
given by 
\EQ
\Theta(x) = 2 \pi g_i (2 - 2 \Delta_i) \,\varphi_i \,\,\,.
\label{trace}
\EN 
It enters a useful sum rule, the so called $\Delta$--theorem 
\cite{DSC} 
\EQ
\Delta_j = -\frac{1}{4 \pi \langle \varphi_j \rangle_i} 
\, \int d^2x \,\langle \Theta(x) \,\varphi_j(0) \rangle_i^c 
\,\,\,, 
\label{Deltath}
\EN 
which relates the conformal dimension $\Delta_j$ of the 
field $\varphi_j$ to its VEV and to the integral of its connected 
off--critical correlator with $\Theta(x)$. It is easy to 
see that the above formula simply expresses the content of the 
fluctuation--dissipation theorem and when the above integral 
diverges, so does the VEV in the denominator, in such a way 
that eq.\,(\ref{Deltath}) always keeps its validity \cite{FMS2}.  

As discussed above, basic quantities entering the universal 
ratios are the generalized susceptibilities $\Gamma_{jk}^i$. 
By using the fluctuation--dissipation theorem, they are given 
by 
\EQ
\hat\Gamma_{jk}^i = \int d^2 x \langle \varphi_j(x) 
\varphi_k(0)\rangle^{c}_i 
\,\,\,.
\label{fluctdis}
\EN 
By extracting their dependence on the coupling constant $g_i$, 
we have $\hat \Gamma_{jk}^i = 
\Gamma_{jk}^i\, g_i^{\frac{\Delta_j + \Delta_k-1}{1-\Delta_i}}$ 
with 
\EQ
\Gamma_{jk}^i = {\cal C}_i^{2 \Delta_j+2\Delta_k-2} 
\int d\tau \frac{1}{\tau^{2\Delta_j+2\Delta_k}} Q_{jk}(\tau) 
\,\,\,.
\label{finalgamma}
\EN
Some of the above susceptibilities can be determined exactly, 
such as the components $\Gamma_{ik}^i$, whose values are 
provided by the $\Delta$--theorem sum rule 
\EQ
\Gamma_{ik}^i = - \frac{\Delta_k}{1-\Delta_k} B_{ki} \,\,\,.
\label{gammachiusa}
\EN 
In all other cases, when an exact formula is not available, 
the strategy to evaluate the generalized susceptibilities  
relies on two different possible representations of the 
correlators. As we will see, these representations have 
the advantage of converging very rapidly in two distinct regions 
and to smoothly join in between. 

The first representation is based on Conformal Perturbation 
Theory (with the use of the non--analytic expression 
of the VEV) \cite{ZamYL}. In this approach the two--point
correlators are expressed as 
\EQ
\langle\varphi_i(x) \varphi_j(0)\rangle = 
\sum_p C_{ij}^p(g;x) \langle A_p(0)\rangle  \,\,\,
\label{correlator}
\EN
where the structure functions 
$C_{jk}^p(g;x)$ admit the 
expansion
\EQ
C_{ij}^p(g;x)= r^{2(\Delta_p-\Delta_i-\Delta_j)} \,
\sum_{n=0}^{\infty}C_{i,j}^{p(n)}(gr^{2-2\Delta_{\Phi}})^n 
\,\,\, ,
\label{Cexpansion}
\EN
($r=\mid x\mid$) and can be computed perturbatively in $g$, the coupling
conjugated to the perturbing operator $\varphi$. 
For instance, the first order correction to the conformal structure 
constant $C_{i,j}^{p(0)}$ is determined by \cite{ZamYL}
\EQ
C_{i,j}^{p(1)} = -\int ' d^2w \,\langle{A}^p(\infty) 
{\varphi}(w){\varphi}_i(1){\varphi}_j(0)\rangle_{CFT} 
\,\,\,,
\label{firstorderc}
\EN
where the prime indicates a suitable infrared (large distance) 
regularization of the integral. This representation allows a 
very efficient estimate of the correlation function in its 
short distance regime $r \ll \xi$. 

The second representation is based on the Form Factors (which 
will be discussed in more detail below) and allows an efficient 
control of the large distance behavior, i.e. when $r \gg \xi$. 
In this second representation, one makes use of the knowledge 
of the off--critical mass spectrum of the theory to express the 
correlators 
as
\EQ
\langle\varphi_i(x) \varphi_j(0)\rangle = \sum_{n=0}^{\infty} 
g_{i,j}^{(n)}(r) \,\,\,,
\label{g_n}
\EN 
where 
\begin{eqnarray}
\label{defff}
&& g_{i,j}^{(n)}(r)  =  \int_{\theta_1 >\theta_2 \ldots>\theta_n} 
\frac{d\theta_1}{2\pi} \dots \frac{d\theta_n}{2\pi}\,
\,e^{-r \sum_{k=1}^n m_k \cosh\theta_k} 
\nonumber \label{spectral} \\
&& 
\times \langle 0|\varphi_i(0)|\dots A_{a_n}(\theta_n)
\rangle  
\langle \dots A_{a_n}(\theta_n)|
\varphi_j(0)|0 \rangle 
\,\,\,. 
\nonumber
\end{eqnarray} 
$|A_{a_1}(\theta_1) \dots A_{a_n}(\theta_n)\rangle$ are the 
multi--particle states relative to the excitations of mass 
$m_k$, with relativistic dispersion relations given by $E 
= m_k \cosh \theta$, $p = m_k \sinh\theta$, where $\theta$ 
is the rapidity variable. The spectral representation 
(\ref{spectral}) is an expansion in the parameter 
$e^{-\frac{r}{\xi}}$, where $\xi^{-1} = m_1$ is the lowest 
mass--gap. 

Basic quantities of the large distance approach are the Form Factors 
(FF), {\it i.e.} the matrix elements of the operators $\varphi_i$ on the 
physical asymptotic states 
\EQ F^{\varphi_i}_{a_1,\ldots
,a_n}(\theta_1,\ldots,\theta_n) \,= \, \langle 0|
\varphi_i(0)|A_{a_1}(\theta_1)\ldots A_{a_n} (\theta_n) \rangle \,\,\,.
\label{form} 
\EN 
%The above quantities are unaffected by renormalization effects
%since physical excitations are present in their definition. 
For scalar operators, relativistic invariance requires that the FF 
only depend on the rapidity differences $\theta_i - \theta_j$. 
In integrable quantum field theories the FF are solutions of some 
functional equations \cite{KW,Smirnov} which will be 
recalled in the next section.  It is worth to stress that for 
most of the operators, it is sufficient to insert into the 
spectral representations only their one--particle and two--particle 
FF in order to reach a very reasonable estimate of the correlators. 

In conclusion, both the expansions (\ref{correlator}) and (\ref{g_n}) 
converge very rapidly in their domain (see, for instance 
\cite{ZamYL,CMpol,DMIMMF}) and therefore it is possible 
to estimate the integral (\ref{fluctdis}) by following this 
procedure: 
\begin{enumerate}
\item Express the integral in polar coordinates as 
\EQ
\hat\Gamma_{jk}^i = 2\pi \int_0^{+\infty} 
d r \,r \,\langle\varphi_j(r) 
\varphi_k(0)\rangle_c^i \,\,\, , 
\EN 
and split the radial integral into two pieces as 
\begin{eqnarray}
& & I  = \int_0^{+\infty}  
d r \,r\,\langle \varphi_j(r) 
\varphi_k(0)\rangle_c^i 
\nonumber \\ 
&& =  \int_0^{R}  
d r \,r\,\langle
\ldots 
\rangle_c^i +  
\int_R^{+\infty}  
d r \,r\,\langle\ldots 
\rangle_c^i \nonumber 
\\
& & \equiv I_1(R) + I_2(R) 
\label{I1I2}
\,\,\,.
\end{eqnarray}
\item 
Use the best available short--distance representation of the 
correlator to evaluate $I_1(R)$ as well as the best available 
estimate of its large--distance representation to evaluate 
$I_2(R)$. 
\item Optimize the choice of the parameter $R$ in such a 
way to obtain the best evaluation of the whole integral. 
In practice, this means looking at that value of $R$ 
for which a plateau is obtained for the sum of $I_1(R)$ 
and $I_2(R)$. Said in another way, $R$ belongs to that 
interval where there is an overlap between the short--distance 
and the long--distance expansion of the correlator. 
\end{enumerate}

\noindent
The above methods have been applied to compute universal 
ratios of several interesting universality classes, such as the 
one of the Ising model \cite{DelfinoURIsing}, the Tricritical Ising 
Model \cite{FMS1,FMS2}, the Self--Avoiding Walks\cite{CMpol}, 
the $q$--state Potts model and percolation \cite{CDperco}, to quote 
few. These field theoretical results have been confirmed through direct 
lattice estimates (see Refs.\,\cite{CTV,CH,SBB,EG}).

\section{Disorder Operator and Fermion-Boson Correspondence}

From the above discussion, we have seen that a great deal of 
information can be extracted for the scaling region of a model 
also in the absence of an exact expression of the correlation 
functions. It is however an important theoretical problem 
to see whether it is possible to obtain a closed and exact formula 
for these quantities. It goes without saying that, in general, 
this is quite a difficult problem, partially solved only for few 
models. In the second part of this talk, we shall discuss a 
particularly interesting example of such a computation, which leads 
to a differential equation ruling the correlators. Our presentation 
is based on the results presented in the paper \cite{GDM}. 

In order to start our discussion, let us consider the (euclidean) action
\EQ
{\cal A}=\int d^2x\,[\partial_\mu\phi^*\partial^\mu\phi+m^2\phi^*\phi
+W(\phi^*,\phi)]\,,
\label{action}
\EN
where the potential $W$ is invariant under the $Z_N$ transformation
$\phi\rightarrow e^{2i\pi/N}\phi$, $\phi^*\rightarrow e^{-2i\pi/N}\phi^*$. 
Let $|0_j\rangle$, $j=0,1,\ldots,N-1$, be the $N$ vacuua of the 
broken phase of this theory, with $\mu_k(x)$, 
$k=1,\ldots,N-1$, the disorder operators which create the 
excitations (kink states) interpolating between the vacua 
$|0_j\rangle$ and $|0_{j+k(mod\,N)}\rangle$. 
These operators carry $k$ units of topological charge 
and 
satisfy the conjugation relation $\mu_k^*(x)=\mu_{N-k}(x)$. The mutual 
non-locality between the order and disorder operators 
gives rise to a phase factor when they go  around each other in the 
euclidean plane ($z=x_1+ix_2$, $\bar{z}=x_1-ix_2$)\,:
\begin{eqnarray*}
\phi(z e^{2i\pi},\bar{z}e^{-2i\pi})\,\mu_k(0,0) &=& e^{\frac{2i\pi k}{N}}
   \phi(z,\bar{z})\,\mu_k(0,0)\,,
\label{ac1}\\
\phi^*(z e^{2i\pi},\bar{z}e^{-2i\pi})\,\mu_k(0,0) &=& 
e^{-\frac{2i\pi k}{N}}
   \phi^*(z,\bar{z})\,\mu_k(0,0)\,.
%\label{ac2}
\end{eqnarray*}
These relations are valid along any $Z_N$-invariant renormalization 
group trajectory flowing out of the phase transition point and 
characterise the operators $\mu_k(x)$ beyond their initial role of 
kink creation operators in the broken phase. The theory (\ref{action}) 
with $W=0$ is a particularly simple example of such a trajectory 
for which, as we will see, the correlation functions of disorder 
operators remain non-trivial due to the non-locality with respect 
to the boson. 

The correlation functions of non-local operators in two-dimensional free 
massive theories were extensively studied from the point of view of the 
isomonodromy theory of differential equations in Ref.\,\cite{SMJ} and a 
series of related papers. Several works (see for instance 
Refs.\cite{BKW,MSS,Alyosha,BB,BL}) have been devoted afterwards 
to dealing with this problem through more direct and general methods of 
quantum field theory. As we have seen above, an important approach to 
compute the correlation functions is the one based on the form factors, 
in which the correlators are expressed as sums over multiparticle 
asymptotic states. In our example this series can be explicitly 
resummed and the correlators are given in terms of solutions 
of non-linear differential equations. Moreover, there is a simple 
unified treatment of the bosonic and fermionic cases, with the final 
result expressed by the correspondence\footnote{We use the notation 
$\tilde{\Phi}(x)\equiv\Phi(x)/\langle\Phi\rangle$.} 
\EQ
\langle\tilde{\mu}_j(x)\tilde{\mu}_k(0)\rangle
=\frac{1}{\langle\tilde{V}_{j/N}(x)\tilde{V}_{k/N}(0)\rangle}\,,
\label{main}
\EN
where the correlators on the l.h.s. are computed in the theory (\ref{action})
with $W=0$, and those on the r.h.s. refer to the operators
$V_\alpha(x)=\exp[i\sqrt{4\pi}\,\alpha\,\varphi(x)]$
in the sine-Gordon theory
\EQ
{\cal A}_{sG}=\int d^2x\left[\frac{1}{2}\,
\partial_\nu\varphi\partial^\nu\varphi-\mu\cos\beta\varphi\right]\,,
\label{sg}
\EN
with $\beta=\sqrt{4\pi}$ and $\mu$ a suitably chosen mass scale. 
The theory (\ref{sg}) with $\beta=\sqrt{4\pi}$ is in 
fact a free fermionic theory \cite{Coleman}. For generic values of $\beta$
the elementary excitations are the solitons and antisolitons interpolating 
between adjacent vacua of the periodic potential. Their interaction is 
attractive as long as $\beta<\sqrt{4\pi}$ and they form topologically 
neutral bound states, the lightest one being the particle interpolated by 
the bosonic field in the action (\ref{sg}). These neutral particles are 
absent from the spectrum of asymptotic states in the repulsive region
$\beta>\sqrt{4\pi}$, and at the point $\beta=\sqrt{4\pi}$ where the 
solitons behave as free Dirac fermions.

Since the solitons are non-local in terms of the field $\varphi(x)$,
the evaluation of both sides of Eq.\,(\ref{main}) amounts to computing
correlation functions of operators which are non-local with respect to 
non-interacting particles, bosons for the l.h.s. and fermions for the r.h.s. 
The different statistics of the two particles is responsible 
for the inversion in Eq.\,(\ref{main}). 

Consider then a theory of free, charge conjugated particles $A$ and 
$\bar{A}$ with mass $m$, and denote by $\Phi_\alpha(x)$ a scalar operator 
having a non-locality phase $e^{2i\pi\alpha}$ ($e^{-2i\pi\alpha}$) with 
respect to (the field which interpolates) the particle $A$ ($\bar{A}$). 
Acting on the vacuum state $|0\rangle$, such an operator produces 
neutral states consisting of pairs $A(\theta)\bar{A}(\beta)$. 
The equations satisfied by the form factors of these theories 
\bea
&& f^\alpha_n(\theta_1,\dots,\theta_n,\beta_1,\dots,\beta_n) = 
\label{ff}\\
&& =\langle 0|\tilde{\Phi}_\alpha(0)|A(\theta_1),\dots,
A(\theta_n),\bar{A}(\beta_1),\dots,\bar{A}(\beta_n)\rangle \nonumber 
\eea
are a particular case of those holding for generic integrable theories 
(see e.g.\,\cite{Alyosha})
\begin{eqnarray*}
&& f_n^\alpha(\theta_1,\dots ,\theta_i ,\theta_{i+1},\dots,\theta_n,\beta_1,
\dots,\beta_n)  = \\
&& =S\,f_n^\alpha(\theta_1,\dots,\theta_{i+1},\theta_i,\dots,\theta_n,\beta_1,
\dots,\beta_n), \\ 
&& f_n^\alpha(\theta_1+2i\pi,\theta_2,\dots,\theta_n,\beta_1,\dots,\beta_n) =\\ 
&& = S\,e^{ 2 i \pi \alpha} 
f_n^\alpha(\theta_1, \dots,\theta_n,\beta_1, \dots,\beta_n), \\ 
&& \textrm{Res}_{\theta_1-\beta_1= i \pi} f_n^\alpha(\theta_1, \dots,\theta_n,\beta_1, 
\dots,\beta_n) = \\
&& = iS^{n-1}(1-e^{ 2 i \pi \alpha})f_{n-1}^\alpha(\theta_2,
..,\theta_n,\beta_2,..,\beta_n),
\end{eqnarray*}
where 
\EQ
S=\left\{
\begin{array}{cl}
1 & \mbox{for free bosons} \\
-1 & \mbox{for free fermions}\,. \\
\end{array}
\right.
\EN
It is easy to check that the solution of the above system of 
equations is given by 
\bea
&& f^\alpha_n(\theta_1,\dots,\theta_n,\beta_1,\dots,\beta_n) =
S^{n(n+2)/2}(-\sin\pi\alpha)^n \nonumber \\
&& \times \,e^{\left(\alpha-\frac12\delta_{S,1}\right)
\sum_{i=1}^n(\theta_i-\beta_i)}\,
\left|A_n\right|_{(S)}\,,
\label{solution}
\eea
where $A_n$ is a $n\times n$ matrix ($A_0\equiv 1$) with entries
\EQ
A_{ij}=\frac{1}{\cosh\frac{\theta_i-\beta_j}{2}}\,,
\EN
and $\left|A_n\right|_{(S)}$ denotes the 
permanent\footnote{The permanent of a matrix differs from the 
determinant by the omission of the alternating sign factors $(-1)^{i+j}$.} 
of $A_n$ for $S=1$ and the determinant of $A_n$ for $S=-1$. 

The two-point correlators are then given by
\begin{eqnarray*}
&& G_{\alpha,\alpha'}^{(S)}(t) =
\langle\tilde{\Phi}_\alpha(x)\tilde{\Phi}_{\alpha'}(0)\rangle=
\sum_{n=0}^\infty\frac{1}{(n!)^2\,(2\pi)^{2n}} \\ 
&& \int d\theta_1\dots d\theta_{n}d\beta_1\dots d\beta_{n} 
\,g_n^{(\alpha,\alpha')}(t\,|\theta_1,\dots,\beta_n)\,,
\end{eqnarray*}
where
\bea
&& g^{(\alpha,\alpha')}_n (t) = 
f_n^\alpha(\theta_1,\dots,\beta_n)f_n^{\alpha'}(\beta_n, \dots,\theta_1)\,
e^{-t e_n} \nonumber \\
&=& (S\sin\pi\alpha\,\sin\pi\alpha')^n\,
e^{(\alpha-\alpha')\sum_{i=1}^n(\theta_i-\beta_i)}\,\label{g}\\
&& \times \, 
\left|A_n\right|_{(S)}^2 \,e^{-t e_n}\,\,\,,
\nonumber 
\eea
and 
\[
t=m|x|\,\,\,\,\,\,\,,
\,\,\,\,\,\,\,\,
e_n=\sum_{k=1}^n (\cosh\theta_k + \cosh\beta_k)\,\,\,.
\]
Analogously to the procedure followed in \cite{BL}, we define
a new $n\times n$ matrix $M_n$ with entries
\bea
&& M_{ij}\equiv M(\theta_i,\beta_j)=
(\sin\pi\alpha\,\sin\pi\alpha')^{1/2}\,
e^{-\frac{t}{2}\cosh\theta_i}\,\nonumber \\
&& \times \frac{h(\theta_i)h^{-1}(\beta_j)}{\cosh\frac{\theta_i-\beta_j}{2}}\,
e^{-\frac{t}{2}\cosh\beta_j}\,\,\,,
\eea
where 
\[
h(\theta)=e^{(\alpha-\alpha')\,\theta/2}\,,
\]
and rewrite $g^{(\alpha,\alpha')}$ as
\EQ
g^{(\alpha,\alpha')}_n=S^n\left|M_n\right|^2_{(S)} =
\left|
\begin{array}{cc}
0 & M_n   \\
M_n^T  & 0 \\
\end{array}
\right|_{(S)}\,.
\EN
Finally we symmetrise with respect to the two sets of rapidities by 
introducing a charge index $\varepsilon$ which is $1$ for a particle $A$ 
and $-1$ for $\bar{A}$, so that we express the correlation functions 
as
\bea 
G_{\alpha,\alpha'}^{(S)}(t) & =& 
\sum_{L=0}^\infty\frac{1}{L!\,(2\pi)^L}\sum_{\epsilon_1\dots\epsilon_L}\int 
d\theta_1\dots d\theta_{L} \nonumber \\
& \times & \left|K_{\epsilon_i \epsilon_j}(\theta_i,\theta_j) 
\right|_{(S)}\,,
\label{fred}
\eea
where we have introduced the $L\times L$ matrices with entries
\bea
&& K_{+-}(\theta,\beta)=M(\theta,\beta)\,,\nonumber\\
&& K_{-+}(\theta,\beta)=M(\beta,\theta)\,, \label{pippo}\\
&& K_{++}(\theta,\beta)=K_{--}(\theta,\beta)=0\,\,\,.\nonumber
\eea
The last equation in (\ref{pippo}) ensures that only the terms with $L=2n$ 
occur in (\ref{fred}). Notice that the dependence on the statistics in 
(\ref{fred}) only reduces to taking the permanent or the determinant of 
the same matrix. According to the theory of Fredholm integral operators 
(see e.g.\,\cite{Schwinger}) the above expression is equal to 
\EQ
G_{\alpha,\alpha'}^{(S)}(t)=\textrm{Det}(1+2\pi\,{\bf K})^{-S}\,,
\EN
so that the correlators of operators having the same non-locality 
phase in the free fermion and free boson theories are the inverse 
of each other. This conclusion was first obtained in \cite{SMJ} 
studying the deformation theory of differential equations.
The result (\ref{main}) follows from the fact that the 
operators $V_\alpha(x)$ have a non-locality phase $e^{2i\pi\alpha}$ 
around the solitons at the sine-Gordon free fermion point (see 
e.g.\,\cite{af}), so that $\mu_j(x)$ and $V_{j/N}(x)$ have the same 
non-locality with respect to the corresponding particles. Since $j$ 
runs between $1$ and $N-1$, we are actually working with 
$0 <\alpha = j/N < 1$.

Once Eq.\,(\ref{main}) has been obtained by using the large 
distance expansion of the correlators, it is an interesting 
check to see how this inversion relation between the two correlators 
works in the short distance limit. Both in the bosonic and fermionic 
case the operators $\Phi_\alpha(x)$ satisfy the OPE 
\EQ
\langle\Phi_\alpha(x)\Phi_{\alpha'}(0)\rangle\sim |x|^{-
\Gamma^{(S)}_{\alpha,\alpha'}}\langle\Phi_{\alpha+\alpha'}\rangle+\dots\,,
\label{ope}
\EN
with
\EQ
\Gamma^{(S)}_{\alpha,\alpha'}=X^{(S)}_{\alpha}+X^{(S)}_{\alpha'}-
X^{(S)}_{\alpha+\alpha'}\,\,\,,
\EN
$X^{(S)}_\alpha$ being the scaling dimensions. In the bosonic case the 
index $\alpha+\alpha'$ is taken modulo $1$.
The scaling dimensions can be computed through the formula 
\cite{DSC}
\EQ
X_\alpha^{(S)}= -\frac{1}{2\pi}\int d^2x
\langle\Theta(x)\tilde{\Phi}_\alpha(0)\rangle_{connected}\,,
\EN
where $\Theta(x)$ is the trace of the energy-momentum tensor. Since the only
non-zero form factor of this operator in the free theories is
\EQ
\langle 0|\Theta(0)|A(\theta)\bar{A}(\beta)\rangle=2\pi m^2\left[-i
\sinh\frac{\theta-\beta}{2}\right]^{\delta_{S,-1}}\,,
\EN
one easily finds
\EQ
X^{(S)}_\alpha=
\left\{
\begin{array}{ll}
\alpha(1-\alpha)\,\,\, , & S=1 \\
\alpha^2\,\,\, , & S=-1 \\
\end{array}
\right.
\EN
in agreement with the conjugation properties $\mu_j^*=\mu_{N-j}$ and 
$V_\alpha^*=V_{-\alpha}$. These results coincide with those of conformal 
field theory with `twist' fields (see \cite{Dixon,AlZam}). It follows
\EQ
\Gamma^{(-)}_{\alpha,\alpha'}=-2\alpha\alpha'\,\,\, ,
\EN
\EQ
\Gamma^{(+)}_{\alpha,\alpha'}=
\left\{
\begin{array}{ll}
2\alpha\alpha'\,, & \alpha+\alpha'<1 \\
2[\alpha\alpha'+1-(\alpha+\alpha')]\,, & \alpha+\alpha'>1\,\,\, . \\
\end{array}
\right.
\EN
The agreement for $1<\alpha+\alpha'<2$ is obtained by observing that in 
this range of $\alpha+\alpha'$, the leading short distance term in the 
fermionic case is not the one in (\ref{ope}) but the first off-critical 
contribution
\EQ
\mu/2\,\int d^2y\langle V_\alpha(x)V_\alpha(0)[V_1(y)+
V_{-1}(y)]\rangle_{\mu=0}\,\,\, ,
\label{off}
\EN
which indeed behaves as $|x|^{2[\alpha\alpha'+1-(\alpha+\alpha')]}$ when 
$x\rightarrow 0$.

In general, the two--point functions we are dealing with can be expressed as 
\EQ
G_{\alpha,\alpha'}^{(S)}(t)=e^{S\,\Upsilon_{\alpha,\alpha'}(t)}\,.
\label{newmain}
\EN
where $\Upsilon_{\alpha,\alpha'}(t)$ is given by 
\cite{SMJ,BL}
\bea
&& \Upsilon_{\alpha,\alpha'}(t)=\frac12\int_{t/2}^\infty\rho d\rho\,\times 
\label{Sigma}
\\
&& \times 
\left[(\partial_\rho\chi)^2-4\sinh^2\chi-
\frac{(\alpha-\alpha')^2}{\rho^2}\tanh\chi\right]\,,\nonumber 
\eea
with $\chi(\rho)$ satisfying the differential equation
\EQ
\partial^2_\rho\chi+\frac{1}{\rho}\,\partial_\rho\chi=2\sinh 2\chi+
\frac{(\alpha-\alpha')^2}{\rho^2}\,\tanh\chi\,(1-\tanh^2\chi) 
\label{diff}
\EN
with suitable boundary conditions.
The short distance behaviour for $\alpha+\alpha'<1$ is given by 
\EQ
\lim_{t\rightarrow 0}G_{\alpha,\alpha'}^{(S)}(t)=\left(C_{\alpha,\alpha'}\,
t^{2\alpha\alpha'}\right)^{-S}\,,
\label{uv}
\EN
with an amplitude that can be deduced from the work of 
Ref.\,\cite{russian1}
on vacuum expectation values in the sine-Gordon model:
\bea
&& C_{\alpha,\alpha'}=2^{-2\alpha\alpha'} \times 
\label{ampl} \\
&& \exp\left\{2\int_0^\infty\frac{dt}{t}
\left[\frac{\sinh\alpha t\,\cosh(\alpha+\alpha')t\,\sinh\alpha't}{\sinh^2t} + 
\nonumber \right. \right.\\
&&\left. \left. -\alpha\alpha'e^{-2t}\right]
\right\}\,.\nonumber 
\eea

Finally, let us discuss the case in which we have a $Z_2$ symmetry. 
This is somehow special since a broken phase with two degenerate 
vacua can be realised in terms of a {\em neutral} boson. In fact, 
a disorder operator with non-locality factor $-1$ is present also 
in the theory of a neutral bosonic free particle. It is easy to check 
that, for $S=1$ and $\alpha=1/2$, the form factors (\ref{solution}) 
imply the factorisation
\EQ
\mu_1=\mu_{(1)}\times\mu_{(2)}\,,\hspace{1cm}N=2
\EN
where $\mu_{(j)}$ is the disorder operator with scaling dimension 
$X^{(+)}_{1/2}/2=1/8$ associated to the neutral boson $A_j$ entering the 
decomposition $A=(A_1+iA_2)/\sqrt{2}$, $\bar{A}=(A_1-iA_2)/\sqrt{2}$.

The fermion-boson correspondence observed above for charged particles has 
an analogue in the neutral case, and the correlation function
\EQ
G(t)=\langle\tilde{\mu}_{(j)}(x)\tilde{\mu}_{(j)}(0)\rangle=
\left[G^{(+)}_{1/2,1/2}(t)\right]^{1/2}
\label{neutral}
\EN
can be related to correlators computed in the theory of a free neutral 
fermion, i.e. in the Ising field theory without magnetic field. To see 
this, let us recall that the correlators of the spin and disorder operators in the 
unbroken phase of the scaling Ising model can be written as \cite{McCoyWu}
\bea
&& \tau_\pm(t)\equiv \langle\tilde{\mu}(x)\tilde{\mu}(0)\rangle\pm
                \langle\sigma(x)\sigma(0)\rangle= \\
&& = \exp\left[\pm\frac12 \chi(t/2)-\frac12 \Upsilon_{1/2,1/2}(t)\right]\,,
\nonumber 
\eea
where the functions $\chi$ and $\Upsilon$ are those of 
Eqs.\,(\ref{Sigma}), (\ref{diff}). Hence, it follows from (\ref{newmain}) 
and (\ref{neutral}) that
\EQ
G(t)=[\tau_+(t)\tau_-(t)]^{-1/2}\,.
\EN
Concerning the short distance behaviour of this correlator, the power law
(\ref{uv}) acquires in this case a logarithmic correction due to the 
`resonance' with the leading off-critical contribution (\ref{off}). Such 
a contribution to the correlator $\langle\tilde{V}_\alpha(x)
\tilde{V}_\alpha(0)\rangle$ behaves as $-C_{\alpha,\alpha}t^{1/2}
[1+(2-4\alpha)\ln t]$ in the limit 
$\alpha\rightarrow 1/2$, so that
\EQ
\lim_{t\rightarrow 0}G_{1/2,1/2}^{(-)}(t)=
\lim_{t\rightarrow 0}G^{-2}(t) = {\cal B}\,t^{1/2}\ln(1/t)\,,
\EN
with
\EQ
{\cal B} = - 4\,\mbox{Res}_{\alpha=1/2}\,C_{\alpha,\alpha}=0.588353..\,\,.
\EN
and this amplitude coincides with that of the product 
$\tau_+(t)\tau_-(t)$ in the Ising model.

The considerations of this section have recently been extended to the case
of massive ghost theories \cite{DMM}. In all these cases the  
correlation functions of disorder operators can be computed exactly for free
theories and are related to the same differential equation 
(\ref{diff}). In absence of interaction, the correlators are non-trivial due
to the non-locality between the operators and the particles. 
The resummation of the spectral series for interacting theories remains a 
challanging task even in presence of integrability.

\section{Non-integrable Deformations}

The free bosonic and fermionic theories discussed above can also be regarded 
as describing phases of spontaneously broken $Z_N$ symmetry. In this dual
vision, the excitations are free kinks $|K_{j,j\pm 1}(\theta)\rangle$ 
interpolating between two adjacent vacua $|0_j\rangle$ and 
$|0_{j\pm 1}\rangle$ (the indices are taken modulo $N$). Let us denote by 
$\Phi_{k/N}$, $k=0,1,\dots,N-1$, the operators we are interested in. They 
correspond to the exponential operators $V_{k/N}$ in the fermionic 
case\footnote{In this case the N degenerate vacua can be identified with 
those of the periodic potential in (\ref{sg}) identified modulo N.}, and 
to the operators $\sigma_k$, dual to the disorder operators $\mu_k$, in 
the bosonic case. These operators create multikink excitations with 
zero topologic charge, i.e. starting and ending in the same vacuum state. 
Form factors on kink states were discussed in \cite{CDperco}. Consider 
for the sake of simplicity the two-kink matrix elements
\EQ
F_{j,k}^\pm(\theta_1-\theta_2)\equiv
\langle 0_j|\Phi_{k/N}(0)|K_{j,j\pm 1}(\theta_1)
K_{j\pm 1,j}(\theta_2)\rangle\,,
\label{ffkinks}
\EN
satisfying the equations 
\begin{eqnarray*}
F_{j,k}^\pm(\theta) & = & S\,F_{j,k}^\mp(-\theta)\,, \\
F_{j,k}^\pm(\theta+2i\pi) &=& F_{j\pm 1,k}^\mp(-\theta)\,, 
\\
\textrm{Res}_{\theta=i\pi}F_{j,k}^\pm(\theta) & = & 
i\,\left[\langle 0_j|\Phi_{k/N}|0_j\rangle + \right. \\
&& \left.  -\langle 0_{j\pm 
1}|\Phi_{k/N}|0_{j\pm 1}
\rangle \right]\,\,\, .
\end{eqnarray*}
Since the generator $\Omega$ of $Z_N$ transformations ($\Omega^N=1$) acts 
on states and operators as ($\omega\equiv e^{2i\pi/N}$)
\bea
&& \Omega\,|K_{j,j+1}(\theta_1)K_{j+1,j+2}(\theta_2)\dots\rangle 
= \nonumber \\
&& = |K_{j+1,j+2}(\theta_1)K_{j+2,j+3}(\theta_2)\dots\rangle\,,
\nonumber \\
&&  \Omega^{-1}\,\Phi_{k/N}(x)\,\Omega  = \omega^k\,\Phi_{k/N}(x)\,,
\nonumber 
\eea
the above form factor equations can be rewritten as 
\bea
F_{j,k}^\pm(\theta+2i\pi) &=& S\,\omega^{\pm k}F_{j,k}^\pm(\theta)\,, 
\nonumber \\
\textrm{Res}_{\theta=i\pi}F_{j,k}^\pm(\theta) & = & 
i\,(1-\omega^{\pm k})\langle 0_j|\Phi_{k/N}|0_j\rangle\,.\nonumber 
%\label{residue}
\eea
Once the identifications $A\longleftrightarrow 
K_{i,i+1}$, $\bar{A}\longleftrightarrow K_{i,i-1}$ are made,
these relations are equivalent to the previous equations for the 
Form Factors with $n=1$. This correspondence is easily extended 
to all values of $n$ and leads to the same form factors 
and correlation functions discussed for the unbroken phase. 

The introduction of a `magnetic field' pointing in the $k$ direction and
breaking explicitely the $Z_N$ symmetry corresponds to adding to the free 
action a term
\EQ
h\int d^2x\,\Psi_k(x)\,,
\EN
where 
\EQ
\Psi_k(x) = \frac{1}{N}\,\sum_{l=0}^{N-1}\omega^{-kl}\,
\Phi_{l/N}(x)\,\,\,. 
\EN
The first 
order corrections to the energy density $\varepsilon_j$ of the vacuum 
state $|0_j\rangle$ and to the mass $m_{j,j\pm 1}$ of the kink 
$K_{j,j\pm 1}$ can be computed according to the Form Factor 
Perturbation Theory proposed in \cite{DMS}
\bea
&& \delta\varepsilon_j\sim h\,\langle 0_j|\Psi_k|0_j\rangle=h\,v\,\delta_{j,k}
\,,\nonumber \\
&& \delta m_{j,j\pm 1}^2\sim h\,\langle 0_j|\Psi_k(0)|K_{j,j\pm 1}(i\pi)
K_{j\pm 1,j}(0)\rangle\,.\nonumber 
\eea
It follows from the residue equation written above that 
\bea
&& \textrm{Res}_{\theta=i\pi}\langle 0_j|\Psi_k(0)|K_{j,j\pm 1}(\theta)
K_{j\pm 1,j}(0)\rangle = \nonumber 
\\
&& \frac{iv}{N}\left(\delta_{j,k}-\delta_{j,k\mp 1}
\right)\,, 
\eea
This formula implies that the correction to the mass of the kinks 
interpolating between the vacua $|0_k\rangle$ and $|0_{k\pm 1}\rangle$ 
is infinite. This divergence simply reflects the fact that these kinks 
become unstable because the magnetic field removes the degeneracy of 
the vacuum $|0_k\rangle$ with the two adjacent vacua. These kinks 
are then confined in pairs $K_{k,k\pm 1}K_{k\pm 1,k}$ and the 
confinement gives rise to a string of bound states with zero 
topologic charge, as discussed in more details in \cite{msg}.

\acknowledgments
This work was supported by the TMR Network "EUCLID. Integrable models 
and applications: from strings to condensed matter", 
contract number HPRN-CT-2002-00325.

\end{document}